\documentclass[12pt]{aastex}
\usepackage{fixltx2e}
\def \lp{\>\> .}
\def \lc{\>\> ,}
\def \arcsec{\hbox{$^{\prime\prime}$}}
\def \arcmin{\hbox{$^{\prime}$}}

\def \cc{cm$^{-3}$}

\def \kms{km\ s$^{-1}$}
\def \kks{K kms$^{-1}$}
\def \nh2{n_{H_2}}
\def \nh1{n_{HI}}

\def \iu{erg s$^{-1}$cm$^{-2}$sr$^{-1}$}

\def \nh3{NH$_3$}
\def \n2h{N$_2$H$^+$}

\def \hh{H$_2$}

\def \ox18{$^{16}$O$^{18}$O}

\def \cp{C$^+$}
\def \C2{C{\sc II}}
\def \H1{{\sc HI}}

\def\tb{$T^{bg}$}
\def\ts{$T^*$}
\def\np{$n($C$^+)$}
\def\n3{$n_u$}
\def\n1{$n_l$}
\def\gu{$g_u$}
\def\gl {$g_l$}

\def \mic{$\mu$m}

\def \be{\begin{equation}}
\def \ee{   \end{equation}}
\def \bf {\begin{figure}}
\def \ef {   \end{figure}}
\def \bc{\begin{center}}
\def \ec{\end{center}}

\def \lc{\>\> ,}
\def \lp{\>\> .}

\begin{document}
\title{Collisional Excitation of the [\C2] Fine Structure Transition in Interstellar Clouds}
\setcounter{footnote}{0}

\author{Paul F. Goldsmith\altaffilmark{1}, William D. Langer\altaffilmark{1}, Jorge L. Pineda\altaffilmark{1}, and T. Velusamy\altaffilmark{1}}

\altaffiltext{1} {Jet Propulsion Laboratory, California Institute of Technology}


\begin{abstract}

We analyze the collisional excitation of the 158 \mic\ (1900.5 GHz) fine structure transition of ionized carbon in terms of line intensities produced by simple cloud models.  
The single \cp\ fine structure transition is a very important coolant of the atomic interstellar medium and of photon dominated regions in which carbon is partially or completely in ionized form. 
The [\C2] line is widely used as a tracer of star formation in the Milky Way and other galaxies.  
Excitation of the [\C2] fine structure transition can be via collisions with hydrogen molecules, atoms, and electrons.   
Analysis of [\C2] observations is complicated by the fact that it is difficult to determine the optical depth of the line.  
We discuss the excitation of the [\C2] line, deriving analytic results for several limiting cases and carry out numerical solutions using a large velocity gradient model for a more inclusive analysis.  
For antenna temperatures up to 1/3 of the brightness temperature of the gas kinetic temperature, the antenna temperature is linearly proportional to the column density of \cp\ irrespective of the optical depth of the transition.
This is appropriately referred to as the effectively optically thin (EOT) approximation.  
We review the critical densities for excitation of the [\C2] line by various collision partners, briefly analyze \cp\ absorption, and conclude with a discussion of  \cp\ cooling and how the considerations for line intensities affect the behavior of this important coolant of the ISM.


\end{abstract}
\setcounter{footnote}{0}

\section{INTRODUCTION}
\label{intro}
Velocity--resolved spectroscopy of the 158 \mic\ fine structure transition of ionized carbon ([\C2]) is currently emerging as a powerful probe of star formation in the interstellar medium of the Milky Way and other galaxies.  Due to its relatively low ionization potential, singly ionized carbon is the dominant form of the element under a wide variety of conditions.  [\C2] can be observed only from above the Earth's atmosphere due to telluric water vapor absorption, except for sources with substantial red shifts.
High angular resolution (interferometric) observations are not yet available, and velocity-resolved spectroscopy is the best avenue for determining what regions are responsible for observed [\C2] emission features.    
Such data for Galactic sources have been extremely rare,  and only with the HIFI instrument on Herschel \citep{degraauw2010} and the GREAT instrument on SOFIA \citep{heyminck2012} have high spectral resolution observations of [\C2] become available to the astronomical community.  

Unlike the situation for molecular tracers (such as carbon monoxide), for \cp\ emission we do not have the possibility of multiple transitions and many isotopologues to allow determination of the volume density, temperature, and optical depth, and thus obtain a reasonably accurate determination of the column density.  
In analyzing [\C2]  data, one can compare line profiles with those of CO, its isotopologues, HI, and other tracers, but one difficulty in determining the quantity of \cp\ is knowing the optical depth of the single transition that is available.  
In general, the line width of sources is sufficiently large that the transitions of $^{13}$C$^+$ are blended with those of $^{12}$C$^+$ (with the few exceptions discussed in \S\ref{galactic}, but which will undoubtedly become more numerous in the near future). 
The situation for absorption is less ambiguous, although in dense foreground clouds the usual assumption that all of the \cp\ is in the ground state may not be valid (see \S \ref{absorption}).  
The present analysis is intended to show that the line intensities commonly found for 158 \mic\ [\C2] emission are sufficiently weak that the integrated intensity is proportional to the \cp\ column density along the line of sight, irrespective of the optical depth of the line, and hence are in the ``Effectively Optically Thin'' (EOT) limit.

\section{CLOUD MODEL}
\label{model}

A significant fraction of gaseous carbon is in the form of \cp\ under a wide range of astrophysical conditions, including ionized, atomic, and molecular phases of the interstellar medium (ISM).  
Thus, a given line of sight can encompass multiple distinct regions, as is reflected in the analysis of large--scale [\C2] survey data \citep[e.g.][]{langer2010}.
Rather than consider a highly specific multi--component model from the outset, we consider each region as a separate source. 
Our conclusion, that most of the [\C2] emission observed is optically thin or in the EOT limit, means that a emission from a given direction can be simply decomposed into contributions from the several regions along the line of sight.  
We will thus consider a single cloud, but the results can straightforwardly be extended to more complex situations.
The treatment follows that outlined by \cite{penzias1975} and elaborated by \cite{linke1977}, but which was suggested by \cite{goldreich1974}

We assume that any background energy density in the cloud at the frequency of the transition is characterized by a blackbody at temperature \tb.  
For centimeter-- and millimeter--wavelength lines, this can be the cosmic microwave background.  
For shorter--wavelength transitions in the immediate vicinity of massive giant molecular clouds, this can be the radiation field of the warm dust, which although likely optically thin, can be characterized by an equivalent blackbody background temperature that produces the same energy density.
Note that in general this is not the same temperature as the ``background'' intensity measured in an observation of the source, as discussed in \S \ref{2levelproblem} and \S \ref{absorption}.


We adopt the escape probability formalism to treat radiative transfer, based on a spherical cloud with velocity proportional to radius.  However, as the optical depths for [\C2] are quite modest, the emergent intensity will not be sensitive to the details of the velocity field or of the cloud morphology.

For the present analysis we consider a uniform cloud.  
Since \cp\ is found in regions having temperatures ranging from $\leq$ 100 K to $\simeq$ 10$^4$ K, this will be an idealization of a typical observation. 
Nevertheless, it does seem that individual spectral features found in the observations taken as part of the {\it GOT--C$^+$} survey originate primarily in regions of a single type and physical conditions \citep{langer2010, velusamy2010, pineda2010}.  
We thus adopt a single kinetic temperature $T^k$, \cp\ density \np, and collision rate, $C$.
The size of the region is $L$, so that the \cp\ column density is $N$(\cp) = \np$L$.  

The \cp\ ion has only two fine structure levels in the ground electronic state.  
The lower J = 1/2 level has statistical weight \gl\ = 2.  
The upper J = 3/2 level has statistical weight \gu\ = 4, and lies at  equivalent temperature \ts\ = $\Delta E/k$ = 91.25 K above the ground state.
The measured transition frequency is 1900.537 GHz \citep{cooksy1986} corresponding to a transition wavelength of 157.74 \mic.
The present discussion is relevant as well to $^{13}$C$^+$, as long as the hyperfine splitting that is present in this species can be ignored.
The spontaneous decay rate for the single [\C2] 158 \mic\ line is 2.3$\times$10$^{-6}$ $s^{-1}$ \citep{,mendoza1983, tachiev2001, wiese2007}.
In \S\ref{rates} we discuss the critical densities for excitation by collisions with different partners.
In Table \ref{symbols} we collect the nonstandard symbols that are used in this analysis. 

\begin{deluxetable}{cc}

\tablewidth{0pt}
\tablecaption{ \label{symbols}  Nonstandard Symbols Used in Analysis of [\C2] Excitation and Emission}
\tablehead{
\colhead{Symbol} & \colhead{Explanation or Definition} \\}
\startdata
$T^{ex}$	&	Excitation Temperature\\
$T^*$		&	Equivalent Temperature (= $h\nu/k$)\\
$T^{bg}$	&	Temperature Characterizing Background Radiation Field Internal to Cloud\\
$T^{kin}$	&	Kinetic Temperature\\
$T_{A, so}$	&    Antenna Temperature Produced by Background Source\\
$G$			&	Background Term (= $[exp(T^*/T^{bg}) - 1]^{-1}$)\\
$K$			&	Kinetic Temperature Term ( = $exp(T^*/T^{kin})$)\\
$X$			&	Ratio of Downwards Collision Rate to Effective Spontaneous Decay Rate\\
$g$			&	Ratio of Statistical Weights ( = $g_u$/$g_l$)\\    
\enddata
\end{deluxetable}
\clearpage

\section{EXCITATION AND EMISSION CALCULATION}
\label{excitation}

\subsection{General two level problem}
\label{2levelproblem}

The excitation temperature of the transition is defined by the relative populations of the upper and lower levels, $n_u$ and $n_l$, respectively, through the standard relationship
\be
\frac{n_u}{n_l}  = \frac{g_u }{g_l} e^{-T^* / T^{ex}} \lp
\label{tex}
\ee
Due to the wide range of conditions under which it is the dominant form of carbon, collisional excitation of \cp\  by electrons \citep{keenan1986, blum1991, wilson2002}, atomic hydrogen \citep{launay1977, barinovs2005} and by molecular hydrogen \citep{flower1977, flower1988, flower1990} can all be important.
The upwards and downwards rate coefficients (cm$^3$s$^{-1}$) are related by detailed balance
\be
R_{lu}/R_{ul} = (g_u / g_l) e^{-T^* / T^{kin}} \lp
\ee
For a single collision partner, the collision rates are equal to the rate coefficients times the density $n$ of that collision partner, thus  
\be
C_{ul} = R_{ul}n {\rm ~and ~} C_{lu} = R_{lu}n\lp
\ee
For a region with multiple collision partners, the upwards and downwards rates are the sum of the rates produced by each.
Within a single region, the kinetic temperatures of different collision partners would likely be the same, but this assumption is not necessary to calculate the combined collision rates.

The energy density in the cloud at  the frequency of the \cp\ transition is given by
\be
U = (1 - \beta)U(T^{ex}) + \beta U(T^{bg}) \lc
\ee
where $\beta$\ is the photon escape probability.
The background temperature, $T^{bg}$, is defined by the energy density of the background radiation field at the frequency of the transition of interest.  It may be an isotropic radiation field such as the cosmic microwave background or the far--infrared background.  
It might also include a local contribution such as that produced by warm dust in a giant molecular cloud with active star formation.  
For this discussion we assume that the \cp-- emitting region is permeated by this background, and that the same background produces the ``background'' level measured in an observed spectrum.  
The latter latter will not be the case for a cloud that is far from the source of the background, which is critical for understanding \cp\ absorption, discussed in \S \ref{absorption}.


For a spherical cloud with a large velocity gradient of the form v $\propto$ r, the escape probability is given by 

\be
\beta = \frac{1 - e^{-\tau}}{\tau} \lc
\ee
where $\tau$ is the peak optical depth of the transition.

Radiative processes include spontaneous emission (rate $A_{ul}$ $s^{-1}$), stimulated emission (rate $B_{ul}U$), and stimulated absorption (rate $B_{lu}U$).  The stimulated rate coefficients are again related by detailed balance through
\be
B_{lu} = (g_u / g_l)B_{ul} \lp
\ee
From the relationship between the stimulated and spontaneous downwards rates,
\be
B_{ul}U = \frac{(1 - \beta) A_{ul}}{e^{T^*/T^{ex}} - 1} + \frac{\beta A_{ul}}{e^{T^*/T^{bg}} - 1} \lp
\ee
For convenience in dealing with the background, we define 
\be
G = \frac{1}{e^{T^*/T^{bg}} - 1} \lp
\ee

The rate equation that determines the level populations includes collisional and radiative processes, and is

\be
n_u(A_{ul} + B_{ul}U + C_{ul}) = n_l(B_{lu}U + C_{lu}) \lc
\label{rate}
\ee

With the preceding equations in this section, and substituting into equation \ref{tex}, we can use equation \ref{rate} to obtain an expression for the excitation temperature
\be
e^{T^* / T^{ex}} = \frac{C_{ul} + \beta(1 + G)A_{ul}}{G\beta A_{ul} + C_{ul}e^{-T^* /T^{kin}}} \lp
\ee
Defining
\be
X = \frac{C_{ul}}{\beta A_{ul}}
\label{X}
\ee
and
\be
K = e^{T^*/T^{kin}} \lc
\ee
we obtain
\be
e^{T^* / T^{ex}} = K\frac{X + 1 + G}{X + GK} \lp
\label{tex3}
\ee 
The quantity $\beta A_{ul}$ is an {\it effective} spontaneous decay rate, since it is is only those photons which escape the cloud that contribute to the downwards transition rate.  $X$ is the  ratio of the downwards collision rate to the effective spontaneous decay rate. 
Thus, irrespective of $\beta$ (and thus of $\tau$), for $X$ $\rightarrow$ 0, $T^{ex}$ $\rightarrow$ $T^{bg}$, and for $X$ $\gg$ 1, $T^{ex}$ $\rightarrow$ $T^{kin}$.  

For this two level system, the optical depth can conveniently be described in terms of that which would occur if there were no excitation, i.e. $T^{ex}$ = 0:
\be
\tau_0 = \frac{hB_{lu}N(C^+)}{\delta v} \lp
\label{tau_0}
\ee
where we approximate the line profile function at line center by $\delta v^{-1}$, and $N(C^+)$ is the total column density of \cp.  
The optical depth in general can be written
\be
\tau = \tau_0 \frac{1 - e^{-T^*/T^{ex}}}{1 + (g_u/g_l)e^{-T^*/T^{ex}}} \lp
\ee
We define the ratio of the statistical weights by
\be
g = \frac{g_u}{g_l} \lc
\ee
and using equation \ref{tex3} the optical depth can be expressed as
\be
\tau = \tau_0 \frac{X(K - 1) + K}{X(K + g) + K[1 + G(1 + g)]} \lp
\label{tau1}
\ee



\subsection{Antenna Temperature and Specific Intensity}

If, as discussed in \S \ref{model}, the background radiation density in the cloud is characterized by a temperature $T^{bg}$ that is also responsible for the ``background'' temperature in a spectral line observation, the antenna temperature (above the background) for a uniform source that fills the entire antenna beam pattern is given by
\be
\Delta T_A = (\frac{T^*}{e^{T^*/T^{ex}} - 1} - \frac{T^*}{e^{T^*/T^{bg}} -1})[1 - e^{-\tau}] \lp
\label{ta_gen}
\ee
This will not be the case for clouds for which the background radiation source subtends only a small solid angle, as for clouds arbitrarily located along the line of sight to the source.  This situation needs to be treated somewhat differently, as discussed in \S \ref{absorption}.

The antenna temperature is proportional to the power collected by the telescope per unit bandwidth, which is proportional to the collecting area of the telescope (the effective area $A_e$ in antenna terminology) and to the normalized response convolved with the brightness distribution on the sky.  
For a uniform source that fills the entire antenna beam, the integral becomes the product of the antenna solid angle $\Omega_a$ and the source brightness.  
For a diffraction limited antenna, $A_e \Omega_a$ = $\lambda^2$.   
The result is that the antenna temperature is proportional to the specific  intensity $I_{\nu}$ (W m$^{-2}$ sr$^{-1}$ Hz$^{-1}$), through the relationship
\be
\label{I_nu}
\Delta T_A = \frac{\lambda^2}{2k} \Delta I_{\nu} \lc
\ee
where $\Delta I_{\nu}$ is the specific intensity of the line above that of the background.
For the [\C2] line, the numerical value of the conversion constant in equation \ref{I_nu} is  9$\times$10$^{14}$   for the specific intensity in the units given above.
%
%
Many observations of the [CII] as well as other sub millimeter spectral lines are carried out by low spectral resolution systems which measure only the integrated intensity.  For conditions as described above, there is a simple relationship between the intensity and the antenna temperature integrated over the spectral line in question.  For [\C2] the relationship in generally--used units is
\be
\int \Delta T_A dv (K km s^{-1}) = 1.43\times 10^{5}\Delta I (erg s^{-1} cm^{-2} sr^{-1}) \lp
\label{ta_unresolved}
\ee

\subsection{\label{limiting}Limiting Cases}

With equations \ref{tex3}, \ref{tau1}, and \ref{ta_gen}, we can calculate the observed antenna temperature for arbitrary background and kinetic temperature as well as excitation rate and \cp\ column density.  We now consider some limiting cases that are also of practical interest.
\subsubsection{Optically Thin Emission} 
\label{thin_emis}

\begin{figure}[h]
\begin{center}
\includegraphics[width = 10cm]{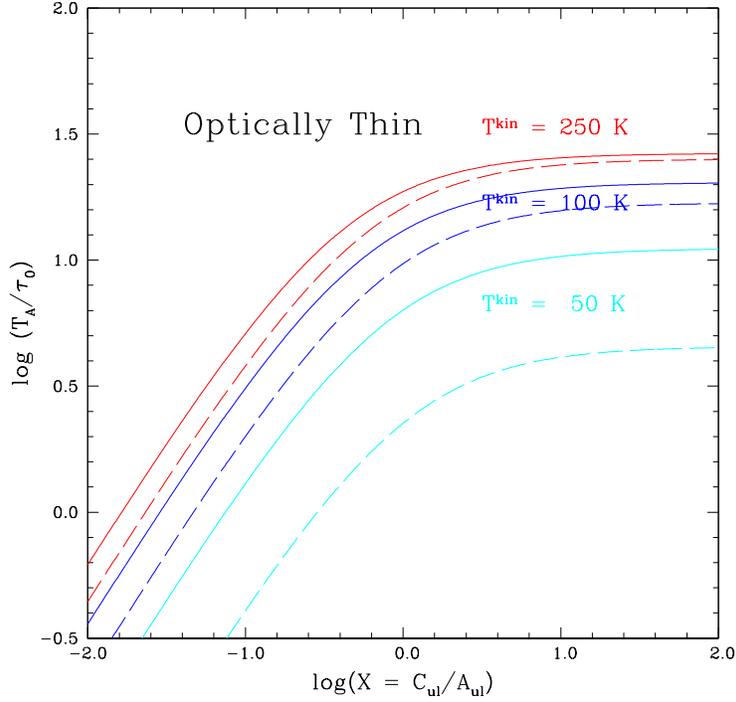}
\caption{\label{fig_thin1} Antenna temperature for optically thin \cp\ emission for three kinetic temperatures.  The antenna temperature is normalized by the no-excitation optical depth $\tau_0$ (equation \ref{tau1}).  The results with no background radiation are indicated by solid lines, while those for a background temperature of 40 K are indicated by dashed lines.  This very large value of the background indicates clearly how the observed antenna temperature (above the background) is reduced by the presence of the background.
}
\end{center}
\end{figure}

For low opacity ($\tau$ $\ll$ 1 and $\beta$ $\simeq$ 1), we have
\be
X = \frac{C_{ul}}{A_{ul}} \lc
\ee
and find that
\be
\Delta T_A (thin) = T^* \frac{\frac{C_{ul}\tau}{A_{ul}}[1 - G(K - 1)]}{\frac{C_{ul}}{A_{ul}}(K-1) + K} \lp
\label{ta_thin}
\ee
Substituting equation \ref{tau1} for the optical depth, we obtain
\be
\Delta T_A (thin) = T^*  \frac{X[1 - G(K - 1)]}{X(g + K) + K[1 + G(1 + g)]}\tau_0   \lc
\label{ta_thin2}
\ee
in which the dependence on the total \cp\ column density and the excitation are clearly separated.  The behavior of the antenna temperature as a function of collision rate (through $X$) for three kinetic temperatures, with and without background, is shown in Figure \ref{fig_thin1}.  
%

In many situations, the background is negligible, and we can simplify the preceding equation to yield
\be
\Delta T_A(thin; no~background) = T^* \frac{ X }{X(g + K) + K  } \tau_0 \lc
\ee
which can also be expressed as
\be
\Delta T_A (thin; no~background) = \frac{ h c^3}{8 \pi k \nu ^2}   \frac{ A_{ul} } {1 + \frac{K}{g} (1 + \frac{A_{ul}}{C_{ul}}) }  {\frac{N(C^+)} {\delta v}}   \lp
\label{ta_thin_no_bg}
\ee
Substituting the relevant quantities for the [\C2] 158 \mic\ line, we obtain
\be
\Delta T_A (thin; no~background) = 3.43\times10^{-16} [ 1 + 0.5e^{91.25/T_{kin}} (1 + \frac{2.4\times 10^{-6}} {C_{ul}})]^{-1} \frac{N(C^+)}{\delta v} \lc
\ee
where the line width is in km s$^{-1}$.
\citet{crawford1985} give essentially equivalent expressions for the specific intensity for the [\cp] line, but not considering any background radiation.  
 
\subsubsection{Optically Thin Subthermal Emission}

For optically thin emission  ($\beta$ $\simeq$ 1), subthermal means that $C_{ul}/A_{ul}$ $<$ 1.  
Equation \ref{ta_thin} in this limit becomes
\be
\Delta T_A(thin; subthermal) = \frac{T^*}{K}\frac{C_{ul}\tau}{A_{ul}}[1 - G(K -1)]\lp
\label{thin_subthermal}
\ee
In this limit, we also find
\be
\tau(subthermal)= \frac{\tau_0}{1 + G(1 + g)} \lc 
\ee 
and we can write
\be
\Delta T_A(thin; subthermal) = \frac{T^* X}{K} \frac{1 - G(K -1)}{1 +G(1 + g)} \tau_0\lp
\label{thin_subthermal2}
\ee
This result can also be derived from equation \ref{ta_thin2} in the low excitation limit, and is convenient, as $\tau_0$ is independent of excitation and is proportional to the total column density of \cp, as seen from equation \ref{tau_0}.   
Equation \ref{thin_subthermal2} reflects the fact that in the low--excitation limit, the population distribution is determined by the background.  The fraction of the total population in the ground state is $(1 + G(1 + g))^{-1}$, and the observed antenna temperature is proportional to the rate of excitation from the ground state.

Substitution of definitions of the radiative and collisional rates allows equation \ref{thin_subthermal2} to be written as
\be
\Delta T_A(thin; subthermal) = \frac{h c^3}{8\pi k \nu^2 \delta v} C_{lu} \frac{1 - G(K -1)}{1 +G(1 + g)} N(C^+) \lc
\ee
where $\delta v$ is the line width in cm s$^{-1}$.  
The value of the deexcitation rate coefficient will depend on the collision partner and may itself be dependent on the kinetic temperature.  
\subsubsection{Optically Thin Thermalized Emission}
\label{thin_thermalized}

At high densities, $C_{ul}$ $\gg$ $A_{ul}$ and in the corresponding limit X $\gg$ 1 we find
\be
\Delta T_A(thin; thermalized) = T^* \frac{1 - G(K - 1)}{g + K}  \tau_0  \lp
\ee
For thermalized optically thin emission, the background reduces the observed antenna temperature by a factor $1 - G(K - 1)$, which is closer to unity than the analogous reduction factor for subthermal excitation.  

\subsubsection{Optically Thick Emission}

For high opacity ($\tau$ $\gg$ 1 and $\beta$ $\simeq$ $\tau^{-1}$ $\ll$ 1), we have
\be
X = \frac{C_{ul}\tau}{A_{ul}} \lc
\ee
and find that
\be
\Delta T_A(thick) = T^* \frac{(\frac{C_{ul}\tau}{A_{ul}}) [1 - G(K - 1)]} {(\frac{C_{ul}\tau}{A_{ul}})(K-1) + K} \lp
\label{ta_thick}
\ee
Equations \ref{ta_thin} and \ref{ta_thick} are very similar but not identical, as the optical depth appears in the denominator for optically thick emission.

\begin{figure}[t!]
\begin{center}
\includegraphics[width = 10cm] {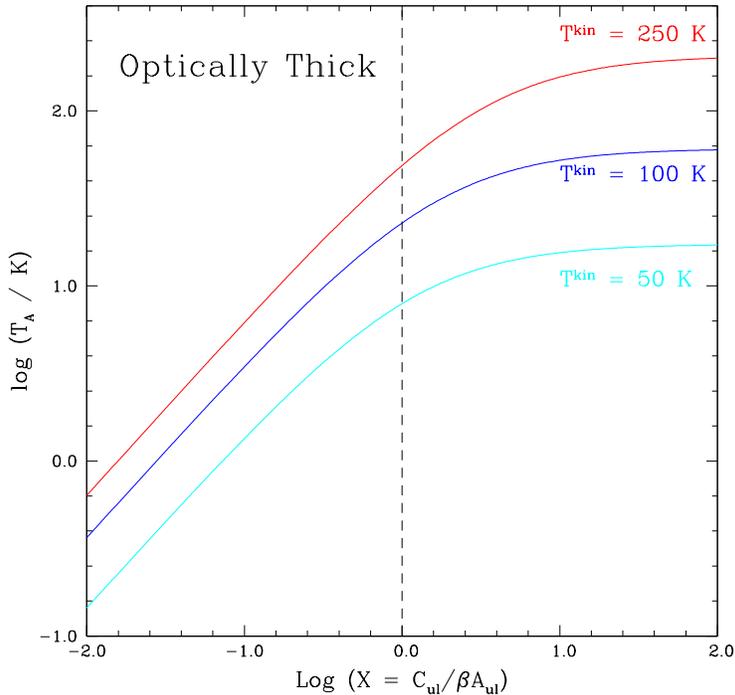}
\caption{\label{fig_thick} Antenna temperature of \cp\ 158\mic\ transition in the optically thick limit as a function of $X$ = $C_{ul}/\beta_{ul}$, for three kinetic temperatures.  In this limit, $\beta$ = $\tau ^{-1}$, so that the antenna temperature is linearly proportional to the optical depth and thus the column density as long as $X \leq$ 1, despite the fact that $\tau \gg$ 1.  There is no background radiation in this example.
}
\end{center}
\end{figure}

The behavior of equation \ref{ta_thick} is illustrated in Figure \ref{fig_thick}, which shows the antenna temperature for the \cp\ line as a function of $X$ = $C_{ul}/\beta A_{ul}$ for three different kinetic temperatures, assuming optically thick emission.  The background is taken to be zero.  The antenna temperature is approximately a linear function of $X$, for $X$ as large as unity, or $C_{ul}$ = $\beta A_{ul}$.
%
%
%
%
%

\begin{figure}[t!]
\begin{center}
\includegraphics[width = 10cm]{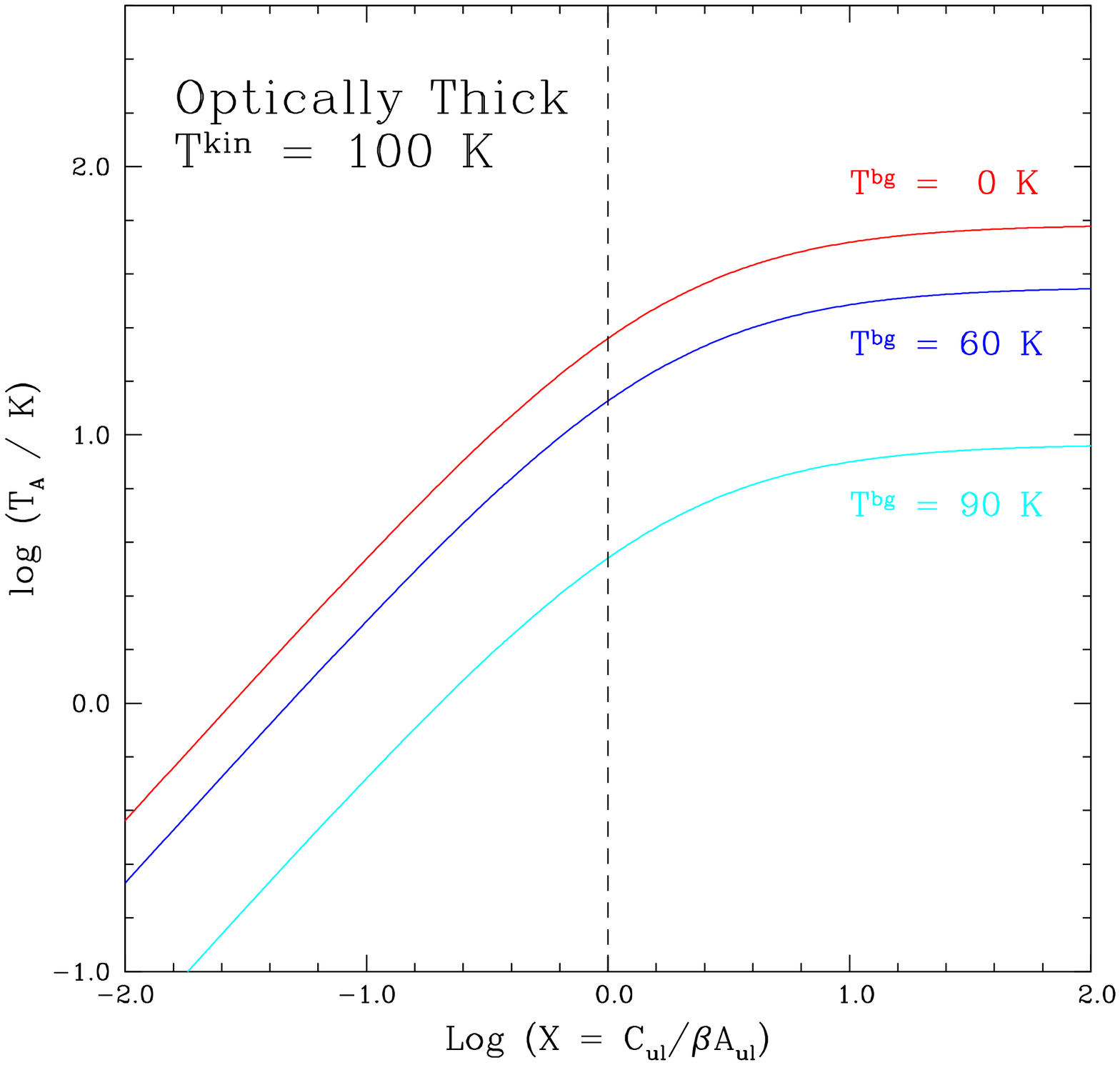}
\caption{\label{bg_effect} Effect of background on the antenna temperature of \cp\ 158 \mic\ transition for optically thick emission.  The background is characterized by a blackbody temperature $T_{bg}$, while the kinetic temperature is 100 K.
}
\end{center}
\end{figure}

As seen in equation \ref{ta_thick}, the effect of the background is to reduce the intensity by a ``background factor" equal to $1 - G(K -1)$, independent of the value of $X$.  This is illustrated in figure \ref{bg_effect}, which shows even for $T^{bg}$ = 60 $K$, the weakening is a modest factor of about 2, while for $T^{bg}$ = 90 $K$, it is a very significant factor of 7.  

\subsubsection{Optically Thick Subthermal Emission}

Subthermal excitation for optically thick emission with $\tau$ $\gg$ 1 means (from equation \ref{tex3}) that $X$ = $C_{ul}/\beta A_{ul}$ = $C_{ul}\tau/A_{ul}$ $\ll$ 1, in which limit equation \ref{ta_thick} becomes 
\be
\Delta T_A(thick;subthermal) = \frac{T^*}{K} \frac{C_{ul}\tau}{A_{ul}} [1 - G(K - 1)]  \lc
\label{thick_subthermal}
\ee
which we see is identical to equation \ref{thin_subthermal}.  Substituting the expression for $\tau_0$ (in the limit $X$ $\ll$ 1), we obtain an expression identical to equation \ref{thin_subthermal2}.  This confirms that in the limit of weak excitation, $X$ $\ll$ 1, the observed intensity is linearly proportional to the total column density even though the optical depth is large.
%
%
\section{General Behavior of the Antenna Temperature}
\label{general}
The preceding discussion has indicated the behavior of the \cp\ fine structure line in a number of limiting cases.  
Unfortunately, the dependence of the excitation on the optical depth and the fact that $\tau$ is itself a function of the degree of excitation are significant obstacles to a general analytical solution.  
We have thus resorted to numerical solution of the rate equations including the effect of radiative trapping and collisional excitation and deexcitation as described in \S\ref{2levelproblem}.  
This allows us to confirm the limits on the conditions under which the antenna temperature is linearly dependent on the \cp\ column density.

We first consider the behavior of $\Delta T_A$ as a function of $\tau_0$.  As given by equation \ref{tau_0}, the zero--excitation optical depth is proportional to the \cp\ column density per unit line width.  For the [\C2] line, this is given by
\be
\tau_0 = 7.49\times 10^{-18} N(C^+)/\delta v (km/s) \lc
\label{tau_0numerical}
\ee
where $\delta v$ is the line width.
Figure \ref{cd_tau0} shows $\Delta T_A$ as a function of $\tau_0$ for a representative situation in which there is no background radiation.
%

\bf[h]
\bc
\includegraphics[width = 10cm]{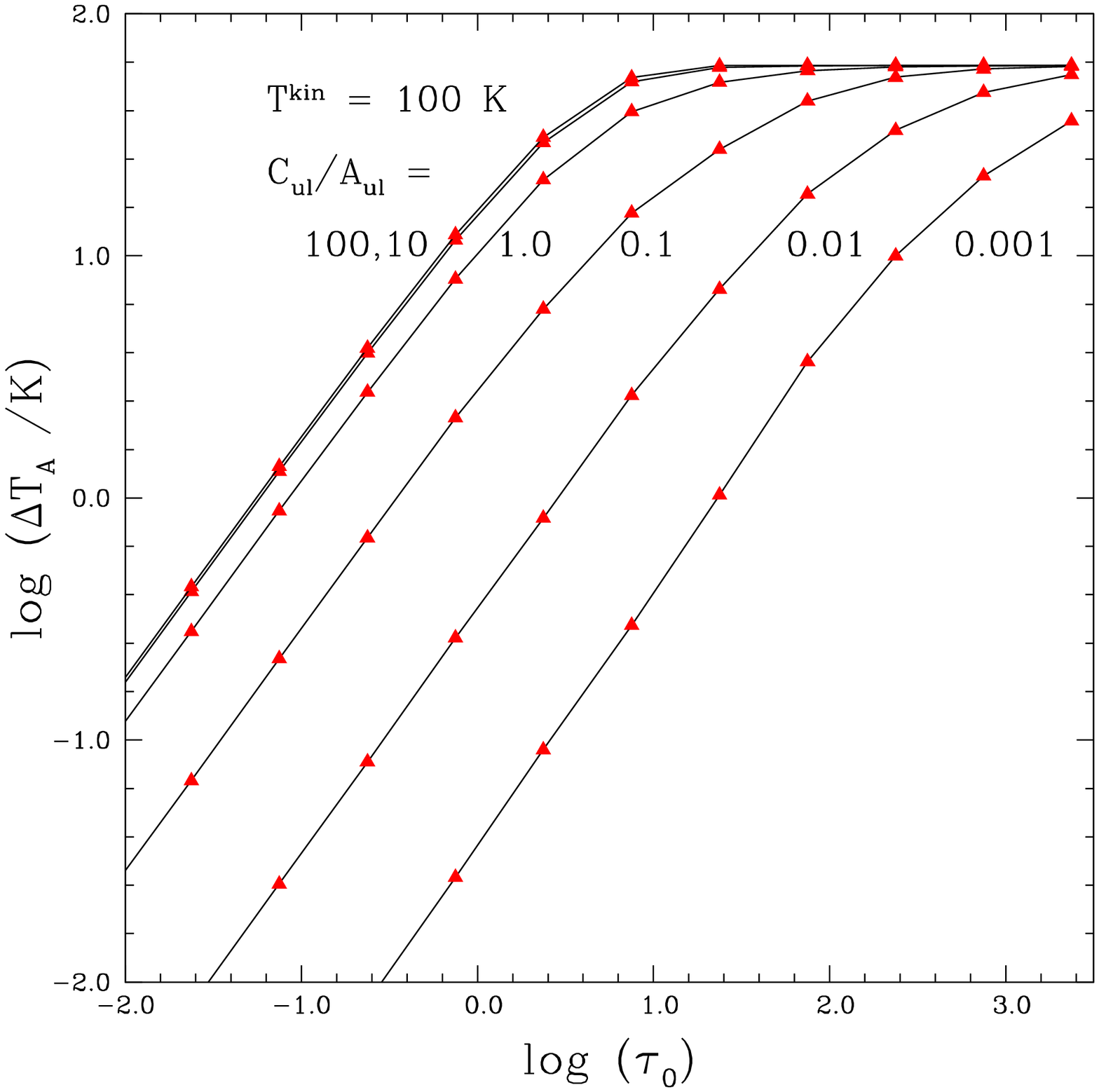}

\caption{\label{cd_tau0} Variation of the antenna temperature of the \cp\ 158 \mic\ transition as a function of the zero--excitation optical depth $\tau_0$, which is proportional to the column density of \cp\ as given by equation \ref{tau_0numerical}.  The kinetic temperature is 100 K, and the different curves are for different ratios of $C_{ul}/A_{ul}$.  
}
\ec
\ef
%
As one approaches thermalization, the antenna temperature becomes independent of the collision rate, since the level populations approach their values characteristic of the kinetic temperature.  The antenna temperature is also limited by the brightness temperature equivalent of the kinetic temperature, which is 61.2 K for the kinetic temperature of 100 K.

Over much of the range of $\tau_0$ the antenna temperature is a linear function of $\tau_0$.  For subthermal excitation, $T_A$ is proportional to $\tau_0$ and to the collision rate, and this situation can apply even to cases in which $\tau_0 \gg\ 1$, as indicated by the preceding discussion.  

From equation \ref{tau1}, for no background
\be
\label{tau_no_bg}
\tau = \tau_0 \frac{X(K - 1) + K}{X(K + g) + K} \lc
\ee
The actual optical depth $\tau$ thus drops from $\tau_0$ in the limit $X\ll 1$ to $\tau_0 (K - 1)/(K + g)$ in the thermalized limit with $X\gg 1$.  
For \cp\ at a kinetic temperature of 100 K, the reduction is a factor $\simeq$ 3, so that the actual optical depth is always within this factor of $\tau_0$.  
It is only as the population of the upper level becomes significant as a result of collisions (possibly enhanced by radiative trapping) that the optical depth starts to differ from $\tau_0$, which also defines the breakdown of the linear relationship between the antenna temperature and the column density.

It is apparent that there are two limits for this linear behavior.  For thermalized emission, the excitation is fixed and the optical depth is a constant times the column density.  For $\tau$ = 0.5, the antenna temperature is $\simeq$ 0.8 times the extrapolated value for optically thin emission.  $\tau_0$ = 1.5 might thus be considered a limit to the linear r\'{e}gime, corresponding to an antenna temperature $\Delta T_A$ = 24 K (for $T^{kin}$ = 100 K).

For non--thermalized emission, the limit is on the quantity $C_{ul}\tau_0/A_{ul}$ rather than just on the optical depth.  
For a fixed value of $C_{ul}\tau_0/A_{ul}$, the deviations from linear behavior increase as $\tau_0$ increases.  
For $C_{ul}/A_{ul}$ $\le$ 1, a limit $C_{ul}\tau_0/A_{ul}$ $\le$ 1 ensures no more than 40\% deviation, and more stringent limit $C_{ul}\tau_0/A_{ul}$ $\le$ 0.5 guarantees linear behavior to within 20\%.

In Figure \ref{cd_tau} we show the dependence of the antenna temperature on the actual optical depth, which is itself computed from the solution of the rate equations.  
While very similar to Figure \ref{cd_tau0}, there are some subtle differences due to the variation of $\tau$ resulting from the changing excitation temperature.  
As discussed above, the transformation from $\tau_0$ to $\tau$ depends on $C_{ul}/A_{ul}$ and on $\beta$.  
This is seen in the deviation of the horizontal positions of the triangles, which are defined by fixed values of $N(C^+)$ and thus $\tau_0$.   
As the excitation increases, the optical depth drops for a given \cp\ column density.  

%
\bf[h]
\bc
\includegraphics[width = 10cm]{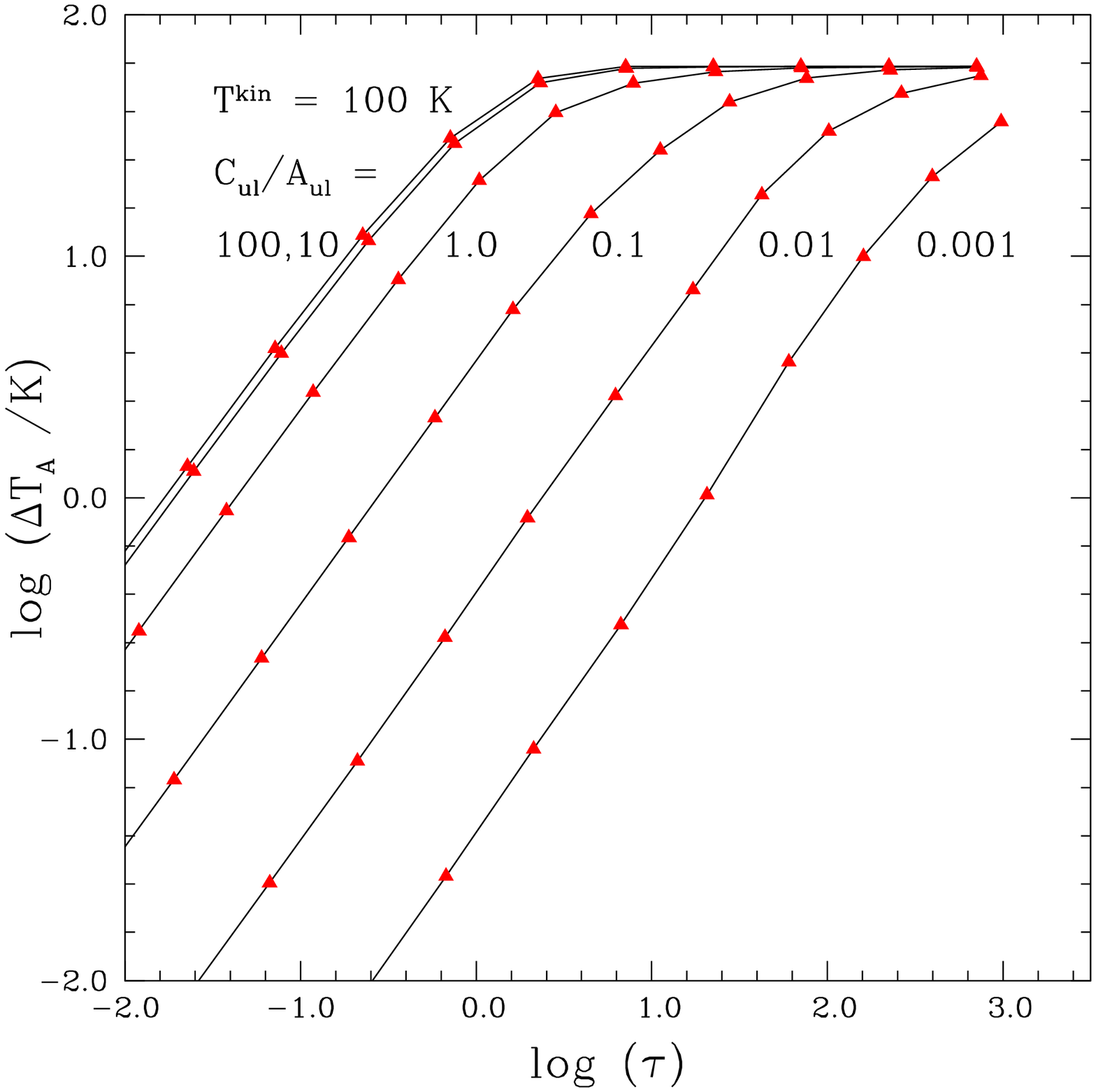}
\caption{\label{cd_tau} Variation of the antenna temperature of the \cp\ 158 \mic\ transition as a function of optical depth $\tau$.  The kinetic temperature is 100 K, and the different curves are for different values of $C_{ul}/A_{ul}$.  
}
\ec
\ef

In the subthermal limit $\tau$ = $\tau_0$, so the antenna temperature is proportional to both of these quantities. 
The demarcation of the limit in which $\Delta T_A$ is proportional to $\tau$ is very similar to that defined by $\tau_0$.  
For thermalized optically thin emission, $\tau$ is a factor of a few lower than $\tau_0$ for reasonable values of the kinetic temperature.  
In consequence, the limit of the linearly proportional relationship is very similar to that derived for $\tau_0$.  
Following the above discussion, we find that for $C_{ul}/A_{ul}$ $\ge$ 10 and $\tau$ $\le$ 0.5 we have a linear relationship to within 20\%.  
For subthermal excitation with $C_{ul}\tau/A_{ul}$ $\le$ 1, the combined limit $C_{ul}\tau/A_{ul}$ $\le$ 1 ensures linear behavior to within 40\%.  
The upper limit on the observed antenna temperature (for $T^{kin}$ = 100 K) is again $\simeq$ 20 K, which more generally corresponds to $\sim$ 1/3 the difference between the brightness temperature of the kinetic temperature minus that of the background temperature.
The nonlinear r\'{e}gime starts at slightly lower values of $\tau$ for lower kinetic temperatures and for significant background.  
A more conservative and broadly applicable limit for sub thermal and thermalized emission is given by $C_{ul}\tau/A_{ul}$ $\leq$ 0.5 is appropriate to broadly ensure proportionality between column density and antenna temperature to within 20\%.  
This can be considered to be the limit of the effectively optically thin (EOT) approximation.

\section{\label{rates} COLLISION RATE COEFFICIENTS AND CRITICAL DENSITIES}

Excitation of the [\C2] transition can be through collisions with electrons, hydrogen atoms, and hydrogen molecules, since \cp\ can be a significant form of carbon in a wide range of conditions in the interstellar medium.   
For each type of collision partner, the deexcitation rate  $C_{ul}$ is the deexcitation rate coefficient $R_{ul}$  multiplied by the collision partner density $n$.  
The critical density $n_{cr}$ is defined by the collisional deexcitation rate being equal to the effective spontaneous decay rate.  
For optically thin emission, the latter is just $A_{ul}$, while more generally it is given by $\beta A_{ul}$.  
We thus have
\be
n_{cr} = \frac{\beta A_{ul}}{R_{ul}} \lp
\label{ncr}
\ee
From equation \ref{X}, this leads to the definition
\be
X = \frac{n}{n_{cr}} \lp
\ee
For optically thin emission with no background radiation, we see from equation \ref{tex3} that
\be
T^{ex}(n_{cr}) = \frac{T^*}{\frac{T^*}{T^{kin}} + ln 2 }\lp
\ee
For \cp\ emission from an atomic hydrogen cloud or warm PDR region, $T^*$ $\simeq$ $T^{kin}$, so that $T^{ex}(n_{cr})$ $\simeq$ 50 K.

Collisions with electrons have been analyzed by many authors \citep{keenan1986, blum1991, blum1992, wilson2002}.  
Over a wide range of temperatures from $\simeq$ 100 K to 20,000 K, the deexcitation collision rate coefficient as a function of the electron temperature can be written
\be
R_{ul} (e^-) = 8.7\times10^{-8} (T^e/2000)^{-0.37} ~~cm^3s^{-1} \lc
\label{electrons}
\ee
where $T^e$ is the electron temperature. 

Collisions between \cp\ and hydrogen atoms have been studied by \citet{launay1977} and more recently in great detail by \citet{barinovs2005}.  The collisional deexcitation rate coefficient as a function of temperature from the most recent analysis is
\be
R_{ul}(H^0) = 4.0\times10^{-11}(16 + 0.35T^{0.5} + 48T^{-1}) ~~cm^3s^{-1} \lp
\ee
A satisfactory fit over a the range 20 K $\leq$ $T^{kin}$ $\leq$ 2000 K is given by
\be
R_{ul}(H^0) = 7.6\times10^{-10} (T^{kin}/100)^{0.14} ~~cm^3s^{-1} \lp
\ee.

The situation for collisions with \hh\ molecules is more complicated due to the existence of two different spin modifications of the molecule, the relatively uncertain ortho-- to--para--ratio, and the rotational states of the molecule that may be populated.
The earliest calculations by \cite{chu1975} were improved upon by \cite{flower1977} and \citet{flower1988}.  
The situation was reviewed by \citet{flower1990}.  
From his cooling rates for \cp\ excited by collisions with \hh\ compared to excitation by collisions with H$^0$, we infer that over a broad range of temperatures, the collision rate coefficients for collisions with molecular hydrogen are approximately half those for collisions with atomic hydrogen.  There are minor differences for the two spin modifications, but the main reason for the smaller rate coefficient for \hh\ is its lower velocity at a given temperature, as the cross sections for \hh\ -- \cp\ collisions are very similar to those for H$^0$ -- \cp\ collisions.  
Over the temperature range appropriate for molecular clouds we thus adopt
\be
R_{ul}(H_2) = 3.8\times10^{-10} (T^{kin}/100)^{0.14} ~~cm^3s^{-1} \lp
\ee
Table \ref{critden} gives the resulting critical densities for different collision partners.  
We see that the critical densities for electron excitation are about a factor of $\simeq$ 50 to $\simeq$ 500  smaller than for excitation by atomic or molecular hydrogen.

The expression $n/n_{cr}$ can be substituted for $X$ in any of the equations in \S\ref{limiting}.  
In general, for densities below the critical density ($X$ $<$ 1) the excitation is sub thermal, and for $n$ $\gg$ $n_{cr}$ ($X$ $\gg$ 1) we achieve thermalization.  
In the event that a single region has multiple collision partners (e.g. \hh\ and H$^0$), the value of $X$ is determined by the sum of the downwards collision rates compared to the spontaneous decay rate.
In practice, while the \cp\ emission from a given line of sight may include contributions from primarily molecular, atomic, or ionized regions, these are likely to be physically distinct, and have quite different kinetic temperatures.  
If the combined emission is optically thin, the total is the sum of that from each region. 
The total emission can still be computed using an expanded version of equation \ref{ta_thin2} with the \cp\ column density and line width, together with the kinetic temperature and background for each region.  

\begin{deluxetable}{cccc}
\tablewidth{0pt}
\tablecaption{\label{critden} Critical Densities for [\C2] 158 \mic\ Fine Structure Line (cm$^{-3}$)}
\tablehead{

\colhead {Temperature} &		&\colhead{Collision Partner}  &\\
\colhead {(K)}                   &\colhead {e$^-$}	&	\colhead{H$^0$}	&	\colhead{H$_2$}\\
}
\startdata
20		&		5	&		3800		&		7600\\
100	&		9	&		3000		&		6100\\
500	&		16	&		2400		&		4800\\
1000	&		20	&		2200		&		4400\\
8000	&		44	&		1600		&		3300\\
\enddata
\end{deluxetable}
\clearpage

\section{Application to Observations}

\subsection{Galactic Emission}
\label{galactic}

New observational facilities are rapidly increasing the quality and quantity of [\C2] observations to the point where making a complete review of [\C2] fine structure emission is not feasible 
We here focus on a very brief overview indicating some of the cases where the previous discussion can be applied and some where conditions make it unlikely to be helpful.  
An obvious concern is that the very simplified model, particularly the assumption of single kinetic temperature and collision rate, means that realistic lines of sight will require more complex, multicomponent models to model accurately the observed emission.  
Nontheless, there appear to be many situations in which general considerations outlined above can be applied.

There have been a number of large--scale surveys of [\C2] in the Galaxy.  
The FIRAS instrument on the {\it COBE} satellite \citep{boggess1992} made essentially all--sky spectrally--unresolved measurements of the [\C2] line with 7$^o$ angular resolution.  
As reported by \citet{fixsen1999}, the intensity varied from a maximum of $\simeq$ 150 \kks\ in the inner galaxy, to $\simeq$ 40 \kks\ in the outer galaxy.  
With reasonable line widths, the peak equivalent antenna temperature is $\leq$ 10 K.  
Thus, the effectively optically thin (EOT) approximation should be applicable, subject to the significant uncertainty and variation in the range of conditions and possible severe beam dilution of small sources in the very large beam.  

\cite{nakagawa1998} used a balloon--borne instrument having a 20 cm effective telescope diameter producing a beam size of 12\arcmin.4 (FHWM) and a Fabry--Perot spectrometer yielding a velocity resolution of 175 \kms.  The survey covered a large fraction of the inner Milky Way but without spectrally resolving the \cp\ emission.  Typical strong peaks given in their Table 2 have intensities I $\simeq$ 10$^{-3}$ \iu.  From equation \ref{ta_unresolved} this is equivalent to a peak antenna temperature of 14 K assuming a line width of 10 \kms.  This suggests that the effectively optically thin approximation is satisfactory and that the observed intensity is proportional to the column density of \cp.

\begin{figure}[h]
\begin{center}
\includegraphics[width = 16cm]{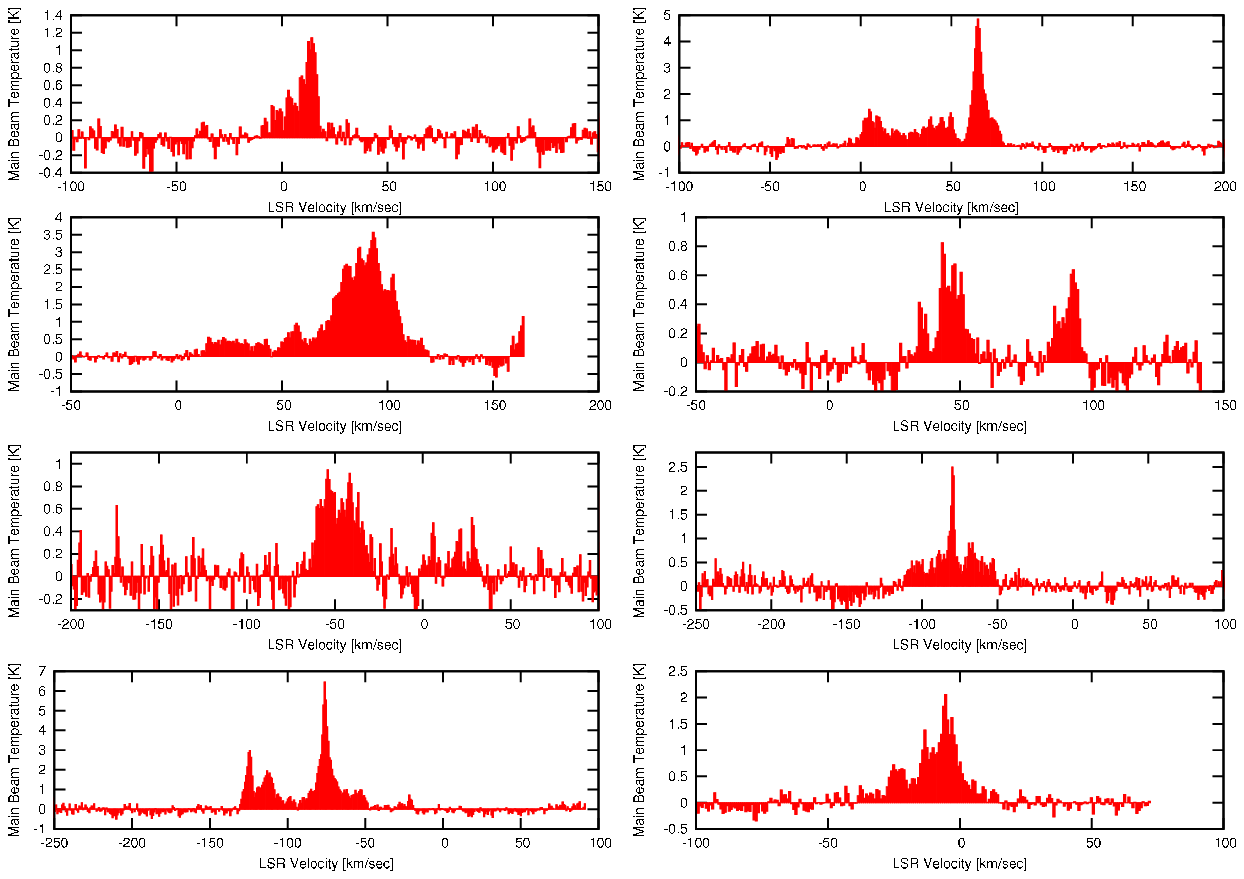}
\caption{\label{CII_spectra} Selection of [\C2] spectra from the {\it GOT-C$^+$} survey.  These spectra were taken at quasi--uniformly spaced directions in the Galactic Plane.  The vertical axis is the antenna temperature corrected for the {\it Herschel} main beam efficiency.
}
\end{center}
\end{figure}

\begin{figure}[h!]
\begin{center}
\includegraphics[width = 12cm]{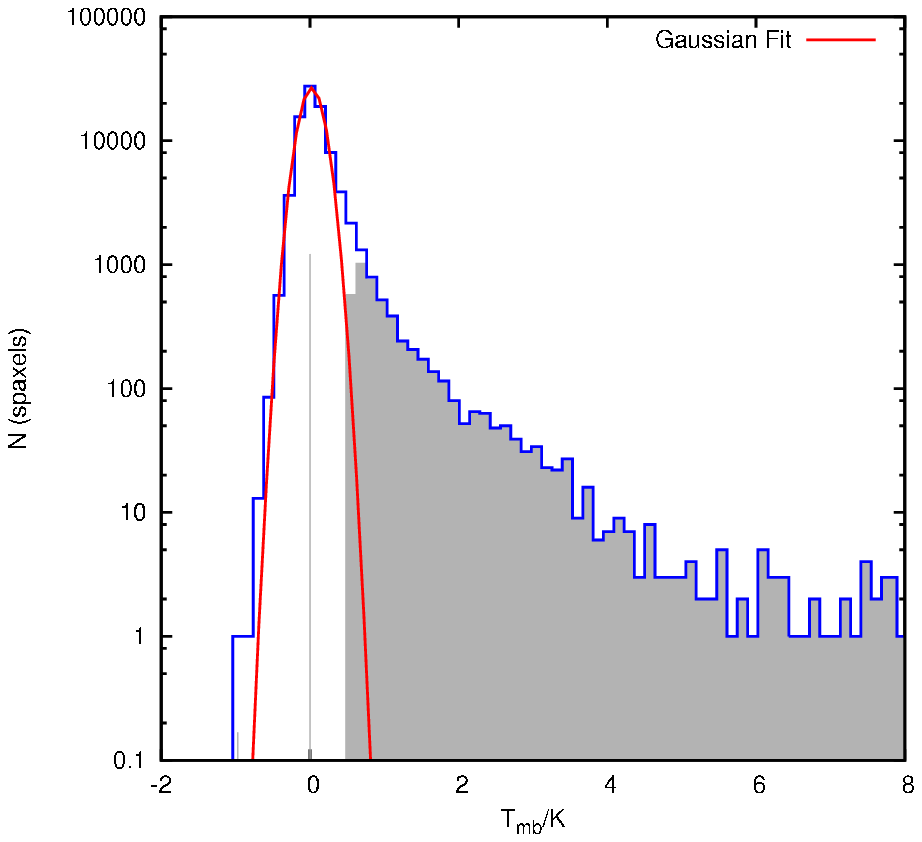}
\caption{\label{CII_histogram} Histogram of main beam temperatures from the {\it Herschel} {\it GOT-C$^+$} survey.  The obvious peak, centered around zero K, is due to the radiometric noise from the HIFI heterodyne instrument.  The distribution of the noise is quite well represented by the Gaussian shown (which was fitted for negative values of the main beam temperature).  The [\C2] emission, indicated by the shaded region, is the excess above this noise, and is evident for $T_{mb}$ between $\simeq$ 0.5 K and 8 K.  Very few spaxels are present with $T_{mb}$ $>$ 8 K.
}
\end{center}
\end{figure}

The availability of sub--\kms\ resolution with the heterodyne instrument HIFI \citep{degraauw2010} on {\it Herschel} \citep{pilbratt2010}, together with the 12\arcsec\ angular resolution afforded by its 3.5 m diameter telescope, enabled the {\it GOT-C$^+$ } Open Time Key Project.  
This survey of the Galactic Plane sampled some 900 lines of sight covering 360$^o$ in longitude, but not targeting specific regions and, of course, being highly undersampled in angle.  
The limited data released to date \citep{langer2010, pineda2010, velusamy2010, velusamy2012} reveal complex spectra with multiple kinematic components.  
Integrated intensities are less than 10 \kks, and peak antenna temperatures rarely exceed 3 K.  
The spectra from the  {\it GOT-C$^+$ } project shown in figure \ref{CII_spectra} illustrate the typical weakness of the [\C2] lines observed in the Galactic Plane when no effort has been made to select regions with ongoing star formation.  
Figure \ref{CII_histogram} shows the distribution of peak main beam brightness temperatures from  the {\it GOT-C$^+$} survey; the most commonly--found values of $T_{mb}$, within a few tenths of a K of zero, are due to the radiometer noise.
The low intensity of the [\C2] emission suggests that the EOT approximation can be used to facilitate data analysis.

The situation is quite different if angularly--resolved bright PDR regions associated with massive young stars are observed.  
This was evident from observations made with incoherent detectors that while not fully resolving the line profiles indicated very strong emission and for reasonable values of line width, large antenna temperatures.  
Examples include observations of M17 by \citet{matsuhara1989} with a 3\arcmin.7 beam, who found a peak intensity of 340 \kks.  
For a line width of 10 \kms\ this corresponds to an antenna temperature of 34 K, which is nearing the limit for the effectively optically thin approximation.  If the kinetic temperature in the PDR region is in excess of 150 K, as predicted by some models, the approximation might still be satisfactory. 
Recent observations of resolved PDR regions include S140 \citep{dedes2010} observed with HIFI and Cepheus B \citep{mookerjea2012} observed by the GREAT instrument on SOFIA.  These both yield moderately strong lines, which are close to the limit of the EOT approximation.

Observations of Orion by \cite{crawford1986} yielded a peak intensity of 5.5$\times$10$^{-3}$ \iu, corresponding to a peak antenna temperature of 150 K if the line width is 5 \kms\ (as revealed by subsequent heterodyne observations).  
These authors concluded that the [\C2] emission is optically thick with $\tau$ $\simeq$ 1.  
The 10 \kms\ velocity resolution of the Fabry--Perot spectrometer employed by \cite{stacey1991} did not quite resolve the [\C2] line.  
The peak intensity measured is 4.1$\times$10$^{-3}$ \iu, which for 5 \kms\ line width corresponds to an antenna temperature of 120 K. 
The velocity resolution was sufficient to resolve the F = 1$\rightarrow$0 component of the [$^{13}$\C2] fine structure line and these authors conclude that while the [\C2] optical depth at the emission peak may approach 2, the emission throughout most of the region mapped is optically thin.  

The Orion PDR region is considered to have densities above 10$^5$ \cc, at least an order of magnitude larger than the critical density for collisions with H$^0$ or \hh\ (Table \ref{critden}).  With $C_{ul}/A_{ul}$ $\geq$ 100, the transition will be thermalized.  
The background continuum is equivalent to an antenna temperature of $\leq$ 10 K and has only minor effect on the emergent line intensity.  
The two analyses discussed above are consistent in suggesting that the optical depth is $\simeq$ 1 in this PDR.  
Since we are not sure to be optically thin, the discussion of $\S$\ref{general} is relevant.  
The optical depth can be determined if the kinetic temperature of the emitting region can be determined (e.g. from model of PDR thermal balance) or inferred from the [$^{13}$\C2] line if the isotope abundance ratio is known.

The first heterodyne observations of the [\C2] line were made by \citet{boreiko1988} who measured a line width of 5 \kms\  and a peak antenna temperature of 86 K with a 55\arcsec\ (FWHM) beam.  
While somewhat lower than that inferred from incoherent measurements, the ability to resolve the F = 2$\rightarrow$1 hyperfine component of the [$^{13}$\C2] line indicated an optical depth $\tau$([$^{12}$\C2]) $\simeq$ 5.  
\citet{boreiko1990} observed a number of bright PDR sources with 0.8 \kms\ velocity resolution and 55\arcsec\ (FWHM) beam size, finding peak antenna temperatures between $\simeq$ 30 K and $\simeq$ 70 K.  
If the kinetic temperatures in the PDRs are in excess of 100 K, their high densities and resulting thermalized emission suggests that the optical depth may be modest, but this is not certain.  
This study also suggested that a significant portion of the [\C2] emission may arise in ionized gas, making it more difficult to determine the \cp\ column density.  

The {\it Herschel} Guaranteed Time Project WADI reported velocity--resolved observations of the HII region NGC 3603 MM2 \citep{ossenkopf2011}. 
The $^{12}$C$^+$ fine structure line has a peak antenna temperature of $\simeq$ 20 K.  
Based on observations of the F = 1$\rightarrow$0 component of the $^{13}$C$^+$ transition, the optical depths may be be small or moderate, depending on the $^{13}$C$^+$/$^{12}$C$^+$ abundance ratio.

The GREAT observations of NGC2024 by \citet{graf2012}  show a striking self--reversal in the $^{12}$C$^+$ line.  
This strongly suggests that there is  low excitation, optically thick gas in front of the bright PDR emission, and that the EOT approximation as described above cannot be directly applied.  
The assertion that the relative intensities of the  $^{13}$C$^+$ hyperfine components cannot be fit by any model in which saturation modifies the observed ratio relative to the optically thin ratio may be due to a problem with the assumed relative line strengths \citep{ossenkopf2012a}.  Despite this complication, the data of \citet{graf2012} and of \citet{ossenkopf2012b} dramatically illustrate the enormous potential value of high resolution spectroscopy in determining the optical depth of the [$^{12}$\C2] line and understanding its relationship to luminous PDRs in the Milky Way as well as in external galaxies.


\subsection{Extragalactic Emission}

[\C2] emission has been recognized as an important tracer of star formation throughout the universe, to a significant degree because of its great luminosity, which is typically between 0.1\% to 1\% of the far--infrared luminosity of star--forming galaxies \citep{madden1993, malhotra2001, stacey2010}.  
The distribution of [\C2] within external galaxies has been studied to a limited extent \citep{poglitsch1995, rodriguez2006}, but the capability of the  {\it Herschel} PACS instrument has allowed much improved studies of its distribution  \citep{mookerjea2011}, in part to determine the fraction of the line emission originating in molecular, atomic, and ionized regions.  
This ubiquity of \cp\ with the varying contributions of these different components of the interstellar medium to the total energy radiated by \cp\ in different galaxies makes calibration of the [\C2]  {\it versus} star--formation rate (discussed by \citet{boselli2002} and \cite{delooze2011}) a difficult task.

From the present discussion we can certainly see that emission from dense PDRs associated with massive young stars is likely thermalized and possibly optically thick.  
The emission thus varies as $n^0$, and as $N^{\alpha}$, with 0 $\leq$ $\alpha$ $\leq$ 1, in contrast to the subthermal, optically thin emission from the diffuse atomic component of the interstellar medium, which varies as $n^1$ and $N^1$.  
The discussion of Galactic emission makes it reasonable that a fraction of the [\C2] emission will be optically thick, but this will likely be a small fraction of the area, if not the total intensity, in most sources. 
The problem of the line opacity is distinct from that of possible absorption by dust, which has been suggested as being significant at the wavelength of [\C2] by \cite{papadopoulos2010}.
The fact that there is as good a correlation as is observed between tracers of star formation and [\C2] emission is likely a reflection of the fact that star formation is associated with PDRs, HII regions, and relatively massive atomic clouds. 
Significant variations would not be very surprising, and the situation may only become more complex when studied with higher angular resolution data as will soon be available from ALMA for red--shifted galaxies at z $\geq$ 1.

\subsection{Absorption by Foreground Clouds}
\label{absorption}

The clouds modeled in \S \ref{excitation} are assumed to be permeated at the frequency of the spectral line of interest by a blackbody radiation field characterized by $T^{bg}$.  
As a result, the excitation temperature of the two level system is limited to the range $T^{bg}$ $\leq$ $T^{ex}$ $\leq$ $T^{kin}$, and there can be no absorption unless $T^{kin}$ is less than $T^{bg}$, which is not physically plausible.
The situation is different if we have regions with disparate background radiation fields.  
This can occur if we have a foreground cloud (with no particular background radiation field) along the line of sight to a strong source of continuum and line emission associated with massive star formation in a GMC.  
In this case, the excitation temperature of [\C2] in the foreground cloud can be as low as the temperature of the CMB, and it is certainly plausible that we see this cloud in absorption.  
\citet{falgarone2010} observed \C2\ absorption (line intensity less than that of continuum) towards the Galactic HII region DR21.  
The distinction between the background and foreground clouds is facilitated by the line of sight velocity difference resulting from Galactic rotation and the long path lengths involved.

The {\it Herschel} Guaranteed Time Key Project (GTKP) {\it PRISMAS}  (M. Gerin, P.I.) targets strong continuum sources specifically to probe absorption of many species along the line of sight.  
The GTKP Project {\it HEXOS} (E. Bergin, P.I.) has detected [\C2] absorption towards Sgr B2.
We denote the antenna temperature (referred to a single sideband) produced by the background source  $T_{A, so}$.
Using equation \ref{ta_gen}, we find that to get absorption, $T^{ex} \leq T^*/ln(1 + T^*/T_{A, so})$.  

$T_{A, so}$ varies from $\simeq$ 2 K to $\simeq$ 10 K \citep{gerin2012a, gerin2012b}.
The maximum \cp\ excitation temperature in the foreground cloud to get absorption ranges from $\simeq$ 25 K for $T_{A, so}$ = 2.5 K to $\simeq$ 40 K for $T_{A, so}$ = 10 K.
For the sources with stronger background antenna temperatures, the \cp\ line is generally seen in absorption, since the excitation temperature in the line of sight clouds can well be less than  40 K.

Equation \ref{tex3} allows us to obtain an upper limit on the density in the foreground cloud that is seen in absorption.  
Assuming that the background radiation field in the foreground cloud is zero, we find $T^{ex}= T^*[T^*/T^{kin} + ln(1 + 1/X)]^{-1}$.  
 Taking the kinetic temperature of the foreground cloud to be 100 K and assuming that the [\C2] line is optically thin and excited by collisions with atomic hydrogen, we find from \S\ref{rates} that n(H$^0$) $\leq$ 200 \cc\ for $T_{A, so}$ = 2.5 K and n($H^0$) $\leq$ 1000 \cc\ for $T_{A, so}$ = 10 K.  
 These densities are generally consistent with what is known about these relatively diffuse foreground clouds.
 The non--zero excitation of the \cp\ means that determining the column density from the calculated spectrum is not as straightforward as for absorption lines from species for which all of the population can be assumed to be in the ground state.
Equation \ref{tau_no_bg} can be used to calculate the optical depth if the excitation temperature or density is known, and equation \ref{tau_0numerical} can then be used to derive the \cp\ column density in the foreground cloud.  
 
If the \cp\ line is assumed to be optically thick, we can use the minimum temperature to obtain $T^{ex}$ and then from equation \ref{tex3} obtain $X$.  
An estimate of the optical depth is required to derive the collision rate and thus the density, but this can yield both the density and \cp\ column density of the foreground cloud.

\section{\cp\ Cooling}

The cooling rate in erg s$^{-1}$ cm$^{-3}$ from the single \cp\ fine structure transition can be written \citep{goldsmith1978}
\be
\Lambda = (A_{ul} n_u + B_{ul} n_u U(T^{bg}) - B_{lu} n_l U(T^{bg}))\beta h\nu \lc
\label{cool1}
\ee
where the upper and lower level $C^+$ densities, $n_u$ and $n_l$, that result from the combination of collisional and radiative processes, are in cm$^{-3}$.
The first term reflects the spontaneous emission that escapes the cloud, the second term is the radiation produced by stimulated emission by the background radiation field (at temperature $T^{bg}$) that escapes from the cloud, and the third term is the radiation absorbed from the background radiation field.
The cooling rate can be expressed as
\be
\Lambda = n_u A_{ul} \beta h \nu [1 - \frac{e^{(T^*/T^{ex})}  - 1}{e^{(T^*/T^{bg})} - 1}] \lc
\label{cool2}
\ee
with the term in square brackets being the background correction term to the usual expression for the cooling rate.
Using the terminology from \S \ref{2levelproblem}, we can write this as
\be
\Lambda = \frac{X[1 - G(K - 1)]}{X(g + K) + K[1 + G(1 + g)]} g \beta A_{ul} n(C^+) h \nu \lc
\label{cool3}
\ee
where $n$(C$^+$) is the total C$^+$ density.
For optically thin emission, $X$ = $C_{ul}/A_{ul}$ and $\beta$ = 1.  
We see that in this limit equation \ref{cool3} becomes
\be
\Lambda = \frac{\frac{C_{ul}}{A_{ul}}[1 - G(K - 1)]}{\frac{C_{ul}}{A_{ul}}(g + K) + K[1 + G(1 + g)]}n(C^+) g A_{ul} h \nu \lp
\label{cool4}
\ee
This form is identical to that of equation \ref{ta_thin2}, except with $g A_{ul} h \nu$ replacing $T^* \tau_0$.  This reflects the fact that for a system with a single transition, the antenna temperature (as well as of course the specific intensity) and the cooling are proportional.
Thus, the previous considerations of thermalization and limits on proportionality between \cp\ column and emission intensity are all applicable.

Equation \ref{cool4} for the case of no background ($G$ = 0) can be written
\be
\Lambda = A_{ul} h \nu n(C^+) [1 + \frac{K}{g} (1 + \frac{A_{ul}}{C_{ul}})]^{-1} \lc
\ee
which is, of course, very similar to equation \ref{ta_thin_no_bg}.  
As an example, we may consider a PDR in which the gas temperature is on the order of 100 K, and the hydrogen is largely molecular.
If the collision rate is much less than the spontaneous decay rate, equivalent to the density being well below the critical density ($\simeq$ 6000 cm$^{-3}$; Table \ref{critden}), the cooling rate is proportional to the C$^+$ density times the \hh\ density.
Adopting the rate for collisions with \hh\ discussed in \S\ref{rates}, we can write the cooling rate as
\be
\Lambda = 9.6\times10^{-24}e^{-91.25/T^{kin}}(\frac{T^{kin}}{100 K})^{0.14} n(H_2)n(C^+) ~erg s^{-1} cm^{-3} \lp
\ee

This expression is the form for optically thin subthermal cooling that is almost always given in treatments of the neutral ISM \citep[e.g.][]{draine2011}.  
More generally, however, especially considering [\C2] line cooling of bright, dense PDRs, we must consider the situation in which the density is high enough that collisional deexcitation is significant, and that the transition may even be thermalized.  
As seen from equation \ref{cool4}, the cooling per \cp\ ion is reduced in this situation.
A similar effect is produced by finite opacity. 
As discussed above, the general case has to be calculated from solution of the nonlinear rate equation, but the results can be obtained by scaling those for the antenna temperature presented in \S\ref{general}.  

\section{CONCLUSIONS}

We have analyzed the radiative transfer and collisional excitation of the [\C2] 158 \mic\ fine structure transition.  
Using an idealized cloud model based on the escape probability approach, we have presented general expressions for the [\C2] antenna temperature that are straightforward to convert to the specific intensity of the fine structure line.
Much of this analysis is applicable to other two level systems, and in may cases to the ground state transition of multilevel systems (such as, for example, the water molecule).
In various relevant limits, analytic expressions can be obtained, which are confirmed by numerical solution of the rate equations including trapping appropriate for spherical cloud large velocity gradient model with $v\propto r$.
In the weak excitation limit defined by $X$ = $C_{ul}/\beta A_{ul}$ $\ll$ 1, the Effectively Optically Thin (EOT) approximation indicates that since every photon produced by spontaneous emission ultimately escapes the cloud, the antenna temperature is proportional to the \cp\ column density even if the optical depth $\tau$ $\gg$ 1.

The strongest lines for which the EOT approximation is valid have $\Delta T_A$ $\leq$ 0.3 times the brightness temperature of the kinetic temperature minus that of the background temperature, or typically 20 K to 30 K for [\C2] line emission from clouds in the ISM.
The [\C2] fine structure line emission from well--resolved PDR regions may be too strong to be in this limit, but interpretation of much of the [\C2] emission traced by large--scale surveys of the Milky Way, and which is likely dominant in a large fraction of the area of external galaxies, will be facilitated by the EOT approximation.
[\C2] cooling from the single fine structure is proportional to the antenna temperature, and consideration of collisional deexcitation and finite optical depth will reduce the cooling per \cp\ ion relative to that obtained in the case of subthermal and  optically thin emission.

\newpage

This work was largely carried out at the Jet Propulsion Laboratory, which is operated by the California Institute of Technology under contract with the National Aeronautics and Space Administration.  It was supported by a Senior Research Scientist leave from JPL in conjunction with a stay as Professeur Invit\'{e} at the \'{E}cole Normale Sup\'{e}rieure, Paris, both of which are gratefully acknowledged.  We thank Maryvonne Gerin for allowing us to use {\it PRISMAS} data prior to publication and for extensive discussions about CII absorption, and Edith Falgarone for interesting discussions about the ISM and the r\^{o}le of ionized carbon.  We greatly appreciate a critical reading of the manuscript by David Neufeld that led to clearing up a significant problem with an earlier version of Figure 1.


\end{document}